\documentclass{fairmeta}

\usepackage[T1]{fontenc}
\usepackage{amsmath,amssymb,amsthm}
\usepackage{newtxtext,newtxmath}
\usepackage{graphicx}
\usepackage{natbib}
\usepackage{booktabs,array,tabularx,multirow}
\usepackage{microtype}
\usepackage{tcolorbox}
\tcbuselibrary{skins,breakable}

\definecolor{toolrcadark}{RGB}{38,94,151}
\usepackage[
  colorlinks=true,
  linkcolor=black,
  citecolor=toolrcadark,
  urlcolor=toolrcadark,
  pdftitle={Rethinking Tool Design for Agentic RCA: A Controlled Empirical Study},
  pdfauthor={Yu Luo, Rongchen Gao, Zhenhui Zhou, Changchang Liu, Yuliang You, Yongqian Sun, Shenglin Zhang, Qiuai Fu, Shijie Wang, Dan Pei}
]{hyperref}
\setcitestyle{authoryear,round,aysep={,},yysep={;}}
\ifdefined\pdfminorversion
\else
  \AtBeginDocument{}
\fi

\newtcolorbox{rcatoolbox}[1]{%
    enhanced,
    breakable,
    title={#1},
    title after break={#1\space{\normalfont\itshape (continued)}},
    lines before break=4,
    fonttitle=\bfseries,
    coltitle=black,
    colback=white,
    colbacktitle=black!7,
    colframe=black!35,
    boxrule=0.4pt,
    arc=2mm,
    outer arc=2mm,
    boxsep=0pt,
    left=6pt,
    right=6pt,
    top=5pt,
    bottom=5pt,
    toptitle=4pt,
    bottomtitle=4pt,
    before skip=7pt,
    after skip=7pt,
    before upper={%
        \setlength{\parindent}{0pt}%
        \setlength{\parskip}{4pt}%
    }
}

\newcounter{findingno}
\newtcolorbox{finding}{
    enhanced,
    colback=white,
    frame hidden,
    borderline west={2pt}{0pt}{gray!60},
    sharp corners,
    left=6pt,
    right=0pt,
    top=1pt,
    bottom=1pt,
    before skip=6pt,
    after skip=6pt,
    before upper={\refstepcounter{findingno}\textbf{Finding~\thefindingno.}\ }
}

\newtcolorbox{rqanswer}[1]{
    colback=gray!6,
    colframe=gray!70,
    boxrule=0.5pt,
    arc=3pt,
    left=5pt,
    right=5pt,
    top=4pt,
    bottom=4pt,
    before upper={\textbf{Answer to RQ#1.}\ }
}

\title{Rethinking Tool Design for Agentic RCA: A Controlled Empirical Study}

\author[1]{Yu Luo}
\author[1]{Rongchen Gao}
\author[1]{Zhenhui Zhou}
\author[1]{Changchang Liu}
\author[1]{Yuliang You}
\author[1,*]{Yongqian Sun}
\author[1]{Shenglin Zhang}
\author[2]{Qiuai Fu}
\author[2]{Shijie Wang}
\author[3]{Dan Pei}

\affiliation[1]{Nankai University}
\affiliation[2]{Huawei Technologies Co., Ltd.}
\affiliation[3]{Tsinghua University}
\contribution[*]{Corresponding Author}

\date{October 2026}
\metadata[Project]{\url{https://luoyu100.github.io/projects/toolrca/project/}}

\abstract{%
Large language model (LLM) agents are increasingly explored for root cause analysis (RCA) in microservice systems, yet empirical guidance on how to design and combine their tools remains limited.
We conduct a controlled empirical study of tool abstraction and composition across models and microservice environments.
We implement 24 structured tools for metric access (L1), evidence analysis (L2), and diagnosis (L3), alongside a Python-based reference setting (L0).
Using 375 failure cases from three microservice systems, we evaluate eight configurations with Qwen3.7-Plus and compare four Qwen models on a shared subset of four configurations.
Our results show that tool configurations affect root cause localization and failure type identification differently.
With Qwen3.7-Plus, L3 achieves 82.8\% top-1 localization accuracy compared with 85.4\% for L0, while requiring less than half the time per case.
Adding tool levels can improve type identification while reducing localization accuracy.
Trajectory analysis reveals cases in which agents override correct diagnostic recommendations after consulting additional evidence.
Model choice also changes tool benefits: adding L3 to L1+L2 improves diagnosis for three models but hurts Qwen3-8B, which rarely invokes L3.
Tool configuration rankings further change across microservice systems.
These findings provide an empirical foundation for designing and using RCA tools, guiding tool selection and composition according to the model, diagnostic objective, and target system.
}

\begin{document}
\maketitle

\begin{figure}[!htbp]
    \centering
    \includegraphics[width=\textwidth]{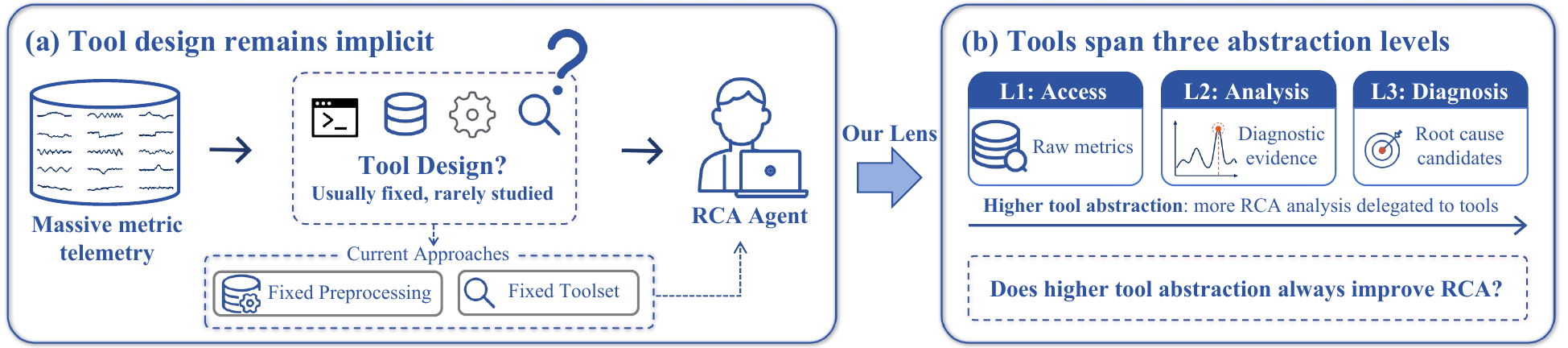}
    \caption{Overview of the motivation, tool abstraction, and key dimensions of our empirical study on tool design for agentic RCA.}
    % Figure description: Overview of existing agentic RCA paradigms, the hierarchy of RCA tool abstraction, and the main dimensions investigated in our empirical study.
    \label{fig:teaser}
\end{figure}

\section{Introduction}
\label{sec:introduction}

Large language model (LLM) agents are rapidly emerging as autonomous problem solvers for complex real world tasks~\citep{yao2023react,schick2023toolformer}.
Root cause analysis (RCA) is a particularly promising application in AIOps, where identifying the entities and failure types behind system incidents requires iterative evidence collection and reasoning for timely mitigation and recovery~\citep{chen2024rcacopilot,roy2024exploring,shetty2024building}.
Recent agentic RCA systems have explored multi-agent collaboration~\citep{zhang-etal-2024-mabc, sun2025trioxpert,luo2025opsagent}, causal and structured reasoning~\citep{jiang2026krca,xie2026foundroot}, SOP guided diagnosis~\citep{pei2025flowofaction,zhao2026oscope,gao2026rcaflow}, explicit reasoning state management~\citep{luo2026graph}, and memory augmented diagnosis~\citep{sun2026opsmem,zhang2026agentic}, substantially advancing automated RCA.
Yet, across these efforts, one fundamental design choice remains largely unexplored: \textbf{\emph{what tools should an RCA agent be given to extract useful information from massive metric streams?}}
This question is critical as metric data in microservice systems is massive and noisy~\citep{zhang2025failure}, while only a small fraction is relevant to the root cause.
Existing systems broadly address this challenge in two ways: some rely on expert designed preprocessing to distill raw metrics into diagnostic signals~\citep{sun2025trioxpert,luo2025opsagent,jiang2026krca}, while others allow agents to actively acquire and analyze metrics using a predefined toolset~\citep{pei2025flowofaction,luo2026graph,sun2026opsmem}.
Despite these efforts, none of these studies has systematically examined what analytical capabilities RCA tools should provide or how those capabilities should be combined.

Through an analysis of how existing agentic RCA systems process metrics, we find that their tools already span different degrees of analytical responsibility.
We capture this hierarchy with three tool levels: \textbf{L1} tools retrieve raw metric data, \textbf{L2} tools transform the data into diagnostic evidence through operations like anomaly detection or causal analysis, and \textbf{L3} tools return diagnostic results. 
OpsAgent's metric processor provides a concrete example of this progression: it retrieves target metric series within a specified time window (L1), applies the 3-sigma rule and temporal shape classification to detect anomalies (L2), and aggregates deviation scores to produce top-5 service instance and metric candidates corresponding to root cause entities and failure types (L3)~\citep{luo2025opsagent}.
As the abstraction level increases, tools encode more domain knowledge and undertake a larger portion of the RCA process, which intuitively suggests that higher level tools should provide stronger support for diagnosis.
\textit{However, does this intuition hold in practice?}

Answering this question requires more than comparing tool configurations in a single setting, because tool effectiveness may depend on both the agent that uses the tool and the environment in which diagnosis is performed.
Models differ in their abilities to analyze data and use tools~\citep{lu-etal-2025-toolsandbox, pmlr-v267-patil25a}, while microservice systems differ substantially in scale, metric space, dependency structure, and failure propagation~\citep{pham2025rcaeval,liang2026metarca}.
We therefore study tool design together with these two contextual factors, conducting controlled experiments across \emph{tool abstraction and composition}, \emph{model capability}, and \emph{microservice system complexity}.
Specifically, we ask three research questions:

\textbf{RQ1: How do tool abstraction and composition affect agentic RCA?}
We examine whether higher-level tools and richer tool combinations consistently improve diagnostic effectiveness and efficiency.

\textbf{RQ2: Does model capability affect the benefits of different tool levels?}
We examine how the performance gains from different tool levels vary across models with different capabilities.

\textbf{RQ3: Does system complexity affect the effectiveness of different tool configurations?}
We study whether diagnostic environments that differ in system scale and complexity favor different tool levels and combinations.

By answering these questions, this paper makes the following contributions:
\begin{itemize}

\item
We derive an abstraction with three levels for structured RCA tools from representative agentic RCA systems and implement 24 structured tools spanning metric access, evidence analysis, and diagnostic methods, establishing a controlled experimental space for studying tool abstraction and composition.

\item
We conduct a controlled empirical study across eight tool configurations, four Qwen models spanning distinct capability levels, and three RCAEval environments with different system architectures and complexity profiles, systematically examining the effects of tool abstraction and composition and their interactions with model capability and system complexity.

\item
We reveal that tool composition can have opposing effects on root cause localization and failure type identification, and that tool benefits depend on model capability and diagnostic environment.
Through trajectory analysis, we identify failures in tool use and conflicts in evidence interpretation that help explain these differences.
Together, these findings provide an empirical foundation for the design and use of tools in agentic RCA.

\end{itemize}
\section{Background and Motivation}
\label{sec:background}

% Define the task and derive the two central design choices.
\textbf{Metric Based Agentic RCA.}
In microservice systems, metric based agentic RCA is complicated by massive and noisy metric spaces along with failure propagation across dependent components, which can separate observed symptoms from their underlying causes~\citep{zhang2025failure,fang2026rethinking}.
For example, a frontend error may stem from \textit{memory exhaustion} in an underlying service or node, illustrating how failure propagation can obscure the true source of an incident and require further evidence across components.
Agentic RCA therefore treats diagnosis as an iterative investigation in which the agent forms hypotheses from available evidence, determines what additional information is needed, acquires and analyzes relevant metrics, then refines its diagnosis until it identifies the root cause entity and failure type~\citep{pei2025flowofaction,luo2026graph,sun2026opsmem}.
Since the complete metric space can easily exceed an LLM's context window, the system must decide \textit{``what evidence to expose to the agent''} and \textit{``how much analysis to delegate to its tools''}.

% Explain why preprocessing can irreversibly restrict available evidence.
\textbf{Challenge 1: Fixed Preprocessing Creates an Evidence Bottleneck.}
Fixed preprocessing uses expert knowledge and statistical procedures to filter, aggregate, and rank metrics into compact diagnostic signals~\citep{sun2025trioxpert,luo2025opsagent,jiang2026krca}.
Although this reduces the search burden and provides salient evidence for diagnosis, it also fixes the evidence boundary before the root cause is known.
If a signal needed to distinguish the true cause is excluded, subsequent reasoning cannot recover it, turning an efficiency mechanism into a potential evidence bottleneck.

\textbf{Challenge 2: Tool Abstraction Creates an Analytical Tradeoff.}
Active metric access relaxes this fixed boundary by allowing agents to retrieve additional metrics as the diagnostic process unfolds~\citep{pei2025flowofaction,luo2026graph,sun2026opsmem}.
However, its effectiveness depends on the abstraction of the tools available to the agent.
Basic tools mainly expose metric data, leaving anomaly detection and evidence association to the agent, whereas higher level tools perform more of this analysis and return processed diagnostic information.
The former preserves greater analytical flexibility but places a heavier burden on the agent, while the latter reduces this burden but increases reliance on embedded analytical methods and may obscure intermediate evidence needed for verification.

% Explain why prior overall system evaluations cannot isolate tool effects.
\textbf{Challenge 3: Lack of Systematic Empirical Guidance.}
Existing agentic RCA methods typically evaluate a predefined toolset as part of a complete diagnostic system~\citep{luo2025opsagent,pei2025flowofaction,luo2026graph,sun2026opsmem}.
Because tool design is coupled with preprocessing, prompts, agent workflows, and model choices, these evaluations cannot isolate its effect on diagnostic performance.
They also provide limited evidence about whether the same configuration remains effective across models with different capabilities or microservice systems with different scale and complexity~\citep{lu-etal-2025-toolsandbox,pmlr-v267-patil25a,pham2025rcaeval,liang2026metarca}.
% Controlled comparisons are therefore needed to determine when different tool levels and combinations are beneficial.

% =========================================================
% Our Study
% Purpose: bridge Background to Section 3.
% Do not repeat the full RQs, findings, or contributions here.
% Simply state how we turn the above challenges into a controlled
% empirical study: tool x model x system.
% =========================================================

\textbf{Our Study.}
We address these challenges through a controlled empirical study centered on RCA tool design.
We construct tools with three different abstraction levels over the same underlying metric data and evaluate both individual levels and their combinations under a consistent agent setting.
We further vary model capability and microservice system complexity to examine how the effectiveness of different tool configurations changes across diagnostic conditions.
By separating tool design from other components of the agentic system, our study provides an empirical basis for understanding what tools should be provided to RCA agents and under what conditions.

\section{Study Design}
\label{sec:study-design}

\begin{figure}[htbp]
    \centering
    \includegraphics[width=\textwidth]{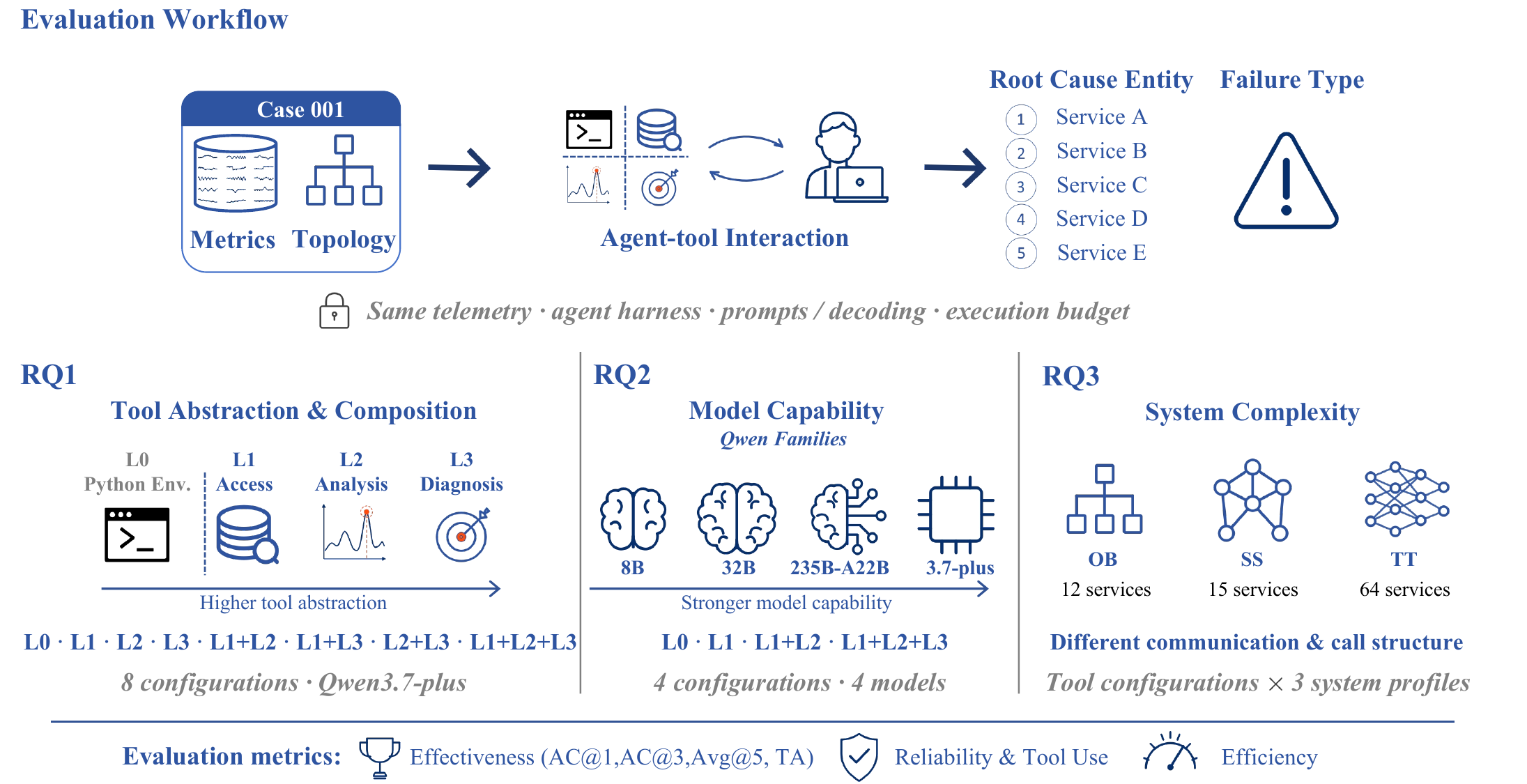}
   \caption{Overview of our empirical study design. The \textbf{top} panel illustrates the evaluation workflow for each failure case. 
   The \textbf{middle} panel summarizes the experimental settings for investigating tool abstraction and composition, model capability, and microservice system complexity, while the \textbf{bottom} panel presents the evaluation dimensions: diagnostic effectiveness, reliability and tool use, and efficiency.}
    % Figure description: Study overview showing case telemetry, iterative agent-tool interaction, diagnostic outputs, the tool, model, and system settings investigated by the three research questions, and the evaluation dimensions.
    \label{fig:study-overview}
\end{figure}

This section presents the design of our empirical study on tool design for agentic RCA.
First, we describe the evaluation workflow used for each failure case.
Next, we introduce the four-level RCA tool spectrum, including the concrete tool implementations and eight experimental configurations.
We then describe the four Qwen models spanning different capability levels and the three RCAEval diagnostic environments with different system architectures and complexity profiles.
Finally, we define the evaluation metrics for diagnostic effectiveness, reliability and tool use, and efficiency.
Figure~\ref{fig:study-overview} summarizes the overall study design and the relationships among these components.

\subsection{Evaluation Workflow}
\label{sec:study-design:workflow}

Each run evaluates a model on a failure case under a specified tool configuration.
We use a ReAct-style agent harness with a shared task prompt and configuration-specific interface instructions.
The task prompt asks the agent to produce a final diagnosis containing ranked root cause candidates and a predicted failure type.
Under common limits on interaction steps and per-response output length, the agent iteratively reasons about the case, invokes an available tool, and uses the returned observation to guide its next action.
A run ends when the agent submits a final diagnosis or reaches the interaction limit.
We then evaluate the diagnostic output against the benchmark ground truth and derive tool-use and efficiency measurements from the execution records, using the metrics defined in Section~\ref{sec:study-design:metrics}.

\subsection{RCA Tool Space and Configurations}
\label{sec:study-design:tools}

\textbf{Tool abstraction.}
By analyzing the tool designs of representative agentic RCA systems~(\textit{e.g.,} OpsAgent~\citep{luo2025opsagent}, Flow-of-Action~\citep{pei2025flowofaction}), we derive a three-level abstraction of RCA tools.
We distinguish these levels by their analytical responsibilities and outputs.
L1 exposes metric observations, basic summaries, and system metadata;
L2 analyzes metrics and returns results such as anomaly scores, temporal features, and metric relationships.
L3 performs diagnosis-oriented analyses that produce root cause candidates or explanations.
Inspired by OpenRCA~\citep{xu2025openrca}, we further introduce L0 as a setting with no predefined RCA tools, where the agent is provided only with the same incident data and a general Python execution interface and must construct the entire analysis process on its own.
L0 serves as a reference for analysis implemented by the agent, while L1--L3 represent increasing abstraction among predefined RCA tools.
Together, they define the tool settings evaluated in this study.

\textbf{Tool implementation.}
We implement a general Python execution interface for L0 and 24 structured RCA tools across L1--L3.
Each tool corresponds to a callable interface, which may support multiple analytical methods.

\begin{rcatoolbox}{L0: Analysis implemented by the agent}

\textbf{Design.}
A general Python interface exposes the case metrics, incident context, and service topology without predefined RCA procedures.
The agent writes and executes code to select and process data, compute statistics, and implement further analyses.

\textbf{Output.}
The interface returns the results produced by the agent's programs as observations.

\textbf{Agent role.}
The agent determines both how to implement the analysis and how to interpret its results.

\end{rcatoolbox}

\begin{rcatoolbox}{L1: Access to metric observations (6 tools)}

\textbf{Design.}
Six interfaces provide metric discovery and retrieval, batch queries, simple aggregation, metric ownership resolution, and local topology inspection.
Metric queries allow the agent to specify a service, metric, and time window.
Batch queries summarize several metrics in one call, while aggregation supports mean, maximum, minimum, sum, and percentiles.

\textbf{Output.}
The interfaces return time series samples, summary statistics, metric ownership information, and local service dependencies.
They do not assign anomaly scores or rank root cause candidates.

\textbf{Agent role.}
The agent chooses what to inspect and interprets the observations to identify anomalies and reason about root causes.

\end{rcatoolbox}

\begin{rcatoolbox}{L2: Evidence analysis (12 tools)}

\textbf{Design.}
Twelve interfaces cover time series processing, anomaly and change point detection, feature extraction, anomalous metric filtering, and relationship analysis.
Anomaly detection supports 3-sigma, IQR, MAD, ESD, and SPOT~\citep{siffer2017anomaly}.
Change point detection includes CUSUM, PELT~\citep{killick2012optimal}, binary segmentation, and Page--Hinkley.
The agent specifies analysis targets and time windows and selects supported methods and parameters.

\textbf{Output.}
Results include anomaly scores, change points, candidate metrics, temporal features, and evidence graphs describing relationships among metrics.

\textbf{Agent role.}
The agent relates these findings across services and determines their diagnostic significance.
An anomaly ranking identifies unusual behavior; translating that evidence into root cause candidates remains the agent's responsibility.

\end{rcatoolbox}

\begin{rcatoolbox}{L3: Diagnostic methods (6 tools)}

\textbf{Design: Candidate generation.}
Propagation analysis searches for paths from specified symptom metrics.
Graph ranking supports PageRank~\citep{brin1998anatomy}, personalized PageRank~\citep{jeh2003scaling}, and random walk with restart~\citep{tong2006fast}.
The classic RCA interface supports MonitorRank~\citep{kim2013root}, CloudRanger~\citep{wang2018cloudranger}, MicroRCA~\citep{wu2020microrca}, Microscope~\citep{lin2018microscope}, MicroCause~\citep{meng2020localizing}, and MicroHECL~\citep{liu2021microhecl}.
It internally filters candidate metrics, prepares their time series, and supplies the selected method with case context and topology.
This interface encapsulates data preparation and diagnostic computation in a single agent action.

\textbf{Design: Evidence integration.}
The fusion interface combines anomaly scores, classic RCA rankings, graph scores, and topology information, and conditionally incorporates change point timing.
It associates these contributions with candidate services and metrics and combines them using predefined scoring rules.
Evidence chain and explanation interfaces gather supporting observations and organize them into diagnostic explanations.

\textbf{Output.}
Depending on the interface, the tools return ranked root cause candidates with scores, propagation paths, or diagnostic explanations with supporting evidence.

\textbf{Agent role.}
The agent decides whether to adopt or revise the diagnostic recommendations and produces the final ranked candidate list and predicted failure type.
Combining L3 with lower-level interfaces allows us to examine whether additional observations and analyses help the agent assess or supplement these recommendations.

\end{rcatoolbox}

L2 and L3 tools access case data and perform their required internal analyses without prior agent calls to lower-level interfaces.
A configuration controls which interfaces the agent can directly invoke, while the underlying data and tool implementations remain unchanged.
Combining levels therefore provides additional analytical entry points without imposing a fixed invocation order.

\textbf{Configuration design.}
We instantiate eight configurations, shown in Table~\ref{tab:tool-configurations}, to isolate the effects of metric access, evidence analysis, diagnostic methods, and their combinations.
L0 captures analysis without predefined RCA tools, while the standalone L1, L2, and L3 configurations evaluate metric access, evidence analysis, and diagnostic methods, respectively.
Because introducing any structured RCA tool changes the defining condition of L0, we evaluate L0 independently and construct compositions only among L1--L3.
The three pairwise combinations examine complementarity between tool levels, while L1+L2+L3 assesses the effect of exposing the full structured toolset.
By comparing configurations that differ by one tool level, we assess the contribution of L3 to lower-level toolsets and the value of adding L1 or L2 when L3 is already available.
Together, these configurations allow us to study individual tool capabilities, complementarity across levels, and the effect of exposing the full structured toolset.

\begin{table}[t]
  \centering
  \caption{Tool configurations and the comparisons enabled by each configuration.}
  \label{tab:tool-configurations}
  \footnotesize
  \setlength{\tabcolsep}{4.0pt}
  \renewcommand{\arraystretch}{1.12}

  \begin{tabular}{@{}p{2.4cm}p{11.2cm}@{}}
    \toprule
    \textbf{Configuration} &
    \textbf{Study Purpose} \\
    \midrule

    L0 &
    Examine diagnosis without structured RCA tools, where the agent performs the analysis on its own. \\

    L1 &
    Isolate the effect of direct access to raw and aggregated metric observations. \\

    L2 &
    Isolate the effect of structured analytical evidence without exposing metric access tools. \\

    L3 &
    Evaluate diagnosis using only diagnostic tools and provide a reference for assessing the value of additional lower-level capabilities. \\

    L1+L2 &
    Examine whether direct metric access and evidence analysis provide complementary benefits. \\

    L1+L3 &
    Examine complementarity between metric access and diagnostic tools through comparisons with L1 and L3. \\
    
    L2+L3 &
    Examine complementarity between evidence analysis and diagnostic tools through comparisons with L2 and L3. \\
    
    L1+L2+L3 &
    Examine whether exposing all three tool levels improves diagnosis over the two-level configurations. \\

    \bottomrule
  \end{tabular}
\end{table}

\subsection{Model Selection Across Capability Levels}
\label{sec:study-design:models}

To examine whether the benefits of different tool levels vary with model capability, we evaluate four Qwen models spanning a broad range of model scale and capability.
We select models from the same Qwen lineage and access them through the same provider, reducing variation introduced by unrelated model families and serving interfaces.
The first three models are drawn from the Qwen3 family~\citep{yang2025qwen3technicalreport}, while Qwen3.7-Plus serves as a stronger proprietary model representing the highest capability level in our study.
Specifically, we evaluate:

\begin{enumerate}
    \item \textbf{Qwen3-8B}, the smallest dense model in our study;
    \item \textbf{Qwen3-32B}, a larger dense model with 32 billion parameters;
    \item \textbf{Qwen3-235B-A22B}, a mixture of experts model with 235 billion total parameters and 22 billion activated parameters; 
    \item \textbf{Qwen3.7-Plus}, which represents the highest capability setting in our study.
\end{enumerate}

We evaluate Qwen3.7-Plus under all eight tool configurations to provide the complete comparison of tool abstraction and composition for RQ1 and the cross-system analysis for RQ3.
For RQ2, we compare all four models under L0, L1, L1+L2, and L1+L2+L3, which span the RCA tool spectrum from no structured RCA tools to the full structured toolset.
This progression allows us to compare how models with different capabilities benefit as tools undertake a larger portion of the RCA process.
The remaining four configurations are evaluated only with Qwen3.7-Plus for the more detailed analyses in RQ1 and RQ3.

\subsection{Datasets and Diagnostic Environments}
\label{sec:study-design:benchmarks}

We evaluate all 375 failure cases in the RE1 suite of RCAEval~\citep{pham2025rcaeval}, an open source benchmark for RCA in microservice systems.
RE1 provides 125 cases for each of Online Boutique (OB), Sock Shop (SS), and Train Ticket (TT).
All three environments cover the same five failure types: CPU, memory, disk, network delay, and packet loss.
Each case includes ground truth labels for the root cause service and injected failure type.
The shared failure taxonomy and equal case counts support comparisons of tool and model effects across the three systems.

Although the three environments follow the same task protocol, their underlying systems differ substantially in service scale, communication structure, and metric dimensionality, as summarized in Table~\ref{tab:diagnostic-environments}.
Rather than reducing system complexity to a single ordered score, we characterize each environment through these observable properties.
OB and SS contain 12 and 15 services, respectively, whereas TT contains 64 services with both synchronous and asynchronous communication and more complex call chains~\citep{pham2025rcaeval}.

\begin{table}[t]
  \centering
  \caption{Profiles of the three diagnostic environments used in our study.}
  \label{tab:diagnostic-environments}
  \footnotesize
  \setlength{\tabcolsep}{4.0pt}
  \renewcommand{\arraystretch}{1.12}

  \begin{tabular}{@{}l c p{5.0cm} c c@{}}
    \toprule
    \textbf{Environment} &
    \textbf{Services} &
    \textbf{Communication and Call Structure} &
    \textbf{Cases} &
    \textbf{Metrics per Case} \\
    \midrule

    Online Boutique (OB) &
    12 &
    Services communicate through gRPC. &
    125 &
    49--59 \\

    Sock Shop (SS) &
    15 &
    Services communicate through HTTP requests. &
    125 &
    57--63 \\

    Train Ticket (TT) &
    64 &
    Services use both synchronous and asynchronous communication
    and form more complex call chains. &
    125 &
    198--238 \\

    \bottomrule
  \end{tabular}
\end{table}

All experiments described above are conducted in each of these three environments, enabling consistent comparisons across systems and allowing us to examine whether the effectiveness of different tool configurations varies across diagnostic environments.

\subsection{Evaluation Metrics}
\label{sec:study-design:metrics}

We evaluate each configuration in terms of diagnostic effectiveness, reliability and tool use, and efficiency.

\textbf{Diagnostic effectiveness.}
We compute all diagnostic accuracy metrics over the full set of evaluated cases, including those for which the agent does not return a valid diagnosis.
For an evaluated set of $N$ cases, let $r_i$ denote the rank of the ground truth root cause entity in the agent's final candidate list for case $i$.
We set $r_i=\infty$ if no valid diagnosis is returned or the ground truth entity is absent from the final candidate list.
We report:
\begin{equation}
    \mathrm{AC@}k =
    \frac{1}{N}\sum_{i=1}^{N}\mathbb{I}(r_i \leq k)\times 100\%,
    \qquad k\in\{1,\ldots,5\}.
\end{equation}
We use AC@1 as the primary localization metric.
We further report
\begin{equation}
    \mathrm{Avg@5}=\frac{1}{5}\sum_{k=1}^{5}\mathrm{AC@}k,
\end{equation}
which summarizes ranking quality within the top five candidates.
For failure type identification, Type Accuracy (TA) is the percentage of evaluated cases for which the agent returns a valid diagnosis whose predicted failure type matches the ground truth.

\textbf{Reliability and tool use.}
Completion measures the percentage of cases in which the agent returns a valid diagnosis within the interaction budget.
We additionally report the L3 usage rate, defined as the percentage of trajectories that invoke at least one L3 tool.

\textbf{Efficiency.}
We report average total token usage, API cost, and end-to-end latency over all evaluated cases.
Token usage includes input and output tokens accumulated across all LLM calls within each run.
\section{Empirical Results}
\label{sec:results}

We organize the empirical results around the three study dimensions introduced in Section~\ref{sec:study-design}.
We first examine how tool abstraction and composition affect RCA under a fixed strong model (RQ1), and then study how these effects change with model capability (RQ2) and system complexity (RQ3).
Within each RQ, we also analyze agent trajectories to explain the observed differences, and say that the agent considers a service when the service appears in its thoughts or tool inputs.
We use AC@1 and TA as the primary metrics for root cause localization and failure type identification, and report AC@3, Avg@5, and efficiency metrics as complementary evidence.

\subsection{RQ1: How do tool abstraction and composition affect agentic RCA?}
\label{sec:results:rq1}

\begin{figure}[ht]
    \centering
    \includegraphics[width=0.95\linewidth]{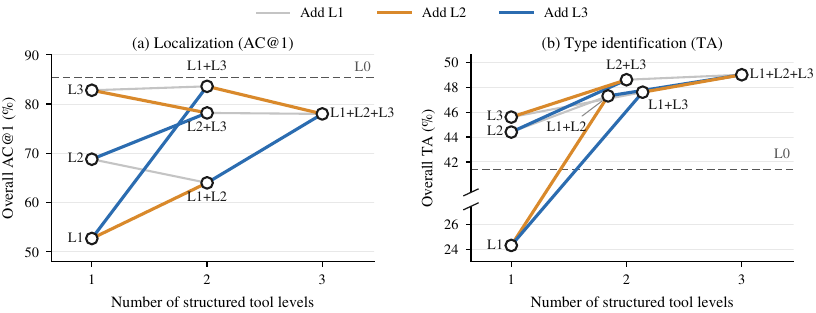}
    \caption{Overall AC@1 (a) and TA (b) of tool configurations under Qwen3.7-Plus. Edges connect configurations that differ by one added tool level; the dashed line marks L0. The TA axis is broken between 27\% and 40\%.}
    \label{fig:rq1-composition}
\end{figure}

We compare all eight configurations under Qwen3.7-Plus, the strongest model in our study.
Table~\ref{tab:rq1_results} reports all metrics, and Figure~\ref{fig:rq1-composition} connects configurations that differ by one tool level.

\begin{table}[t]
\centering
\caption{Overall effectiveness and efficiency of different tool configurations using Qwen3.7-Plus. AC@1, AC@3, Avg@5, and TA are reported as percentages. Tokens are reported in thousands, latency in seconds, and API cost in US dollars, all averaged per case.}
\label{tab:rq1_results}
\resizebox{\columnwidth}{!}{
\begin{tabular}{ll*{7}{c}}
\toprule
\multirow{2}{*}{Seed LLM}
& \multirow{2}{*}{Tool Configuration}
& \multicolumn{4}{c}{Effectiveness}
& \multicolumn{3}{c}{Efficiency} \\
\cmidrule(lr){3-6}
\cmidrule(lr){7-9}
& & AC@1 & AC@3 & Avg@5 & TA
& Tokens & Latency & Cost \\
\midrule
\multirow{8}{*}{Qwen3.7-Plus}
& L0       & \textbf{85.4} & 93.6 & 92.7 & 41.4 & 50.4 & 88.6 & 0.019 \\
& L1       & 52.7 & 70.0 & 69.1 & 24.3 & 53.7 & 51.4 & 0.018 \\
& L2       & 68.8 & 92.3 & 87.5 & 44.4 & 142.9 & 52.8 & 0.044 \\
& L3       & 82.8 & \textbf{96.2} & \textbf{93.7} & 45.6 & \textbf{22.8} & \textbf{35.8} & \textbf{0.008} \\
& L1+L2    & 64.0 & 88.4 & 84.4 & 47.3 & 111.6 & 57.1 & 0.035 \\
& L1+L3    & 83.6 & 94.3 & 92.5 & 47.6 & 45.0 & 39.9 & 0.015 \\
& L2+L3    & 78.2 & 95.6 & 92.3 & 48.6 & 79.0 & 40.0 & 0.025 \\
& L1+L2+L3 & 78.0 & 94.8 & 92.1 & \textbf{49.0} & 95.1 & 48.8 & 0.030 \\
\bottomrule
\end{tabular}
}
\end{table}

\textbf{Higher abstraction helps both targets, but at different levels.}
Among the standalone structured configurations, AC@1 rises from 52.7\% with L1 to 68.8\% with L2 and 82.8\% with L3 (Figure~\ref{fig:rq1-composition}(a)).
L3 comes within 2.6 percentage points of L0, the most accurate configuration at 85.4\%, and exceeds it in AC@3 and Avg@5.
TA instead jumps from 24.3\% with L1 to 44.4\% with L2 and barely moves with L3 (45.6\%; Figure~\ref{fig:rq1-composition}(b)).
L0, which localizes best, identifies failure types worse (41.4\%) than every configuration with L2 or L3.
This ordering does not hold in every system: in AC@1, L2 falls below L1 on Sock Shop and exceeds L3 on Online Boutique, a dependence we examine in RQ3.

\begin{finding}
Higher-level tools improve localization up to L3, which nearly matches L0, but improve type identification mainly from L1 to L2.
L0 localizes best yet trails every configuration with L2 or L3 in TA.
\label{finding:abstraction}
\end{finding}

\textbf{Combining levels rarely helps localization but never hurts type identification.}
In AC@1, a combination rarely improves on its strongest constituent.
L1+L2 (64.0\%) falls below L2 alone, and both L2+L3 (78.2\%) and L1+L2+L3 (78.0\%) fall below L3 alone.
L1+L3 is the only combination that outperforms both of its constituents, yet it improves on L3 by only 0.8 points.
The edges in Figure~\ref{fig:rq1-composition}(a) further show that the effect of a level depends on what it is added to.
Adding L3 improves every configuration that lacks it, by 9.4 to 30.9 points, and adding L1 changes AC@1 by at most 4.8 points in either direction.
Adding L2 improves L1 by 11.3 points but lowers L3 and L1+L3 by 4.6 and 5.6 points, respectively.
TA behaves differently: in Figure~\ref{fig:rq1-composition}(b), every edge points upward, as each of the same nine additions raises TA by 0.4 to 23.3 points, and the full toolset reaches the highest TA of 49.0\%.
Once L2 or L3 is present, however, further levels add at most 4.2 points.

% Effectiveness–efficiency trade-offs 图
\begin{figure}[htbp]
    \centering
    \includegraphics[width=0.9\textwidth]{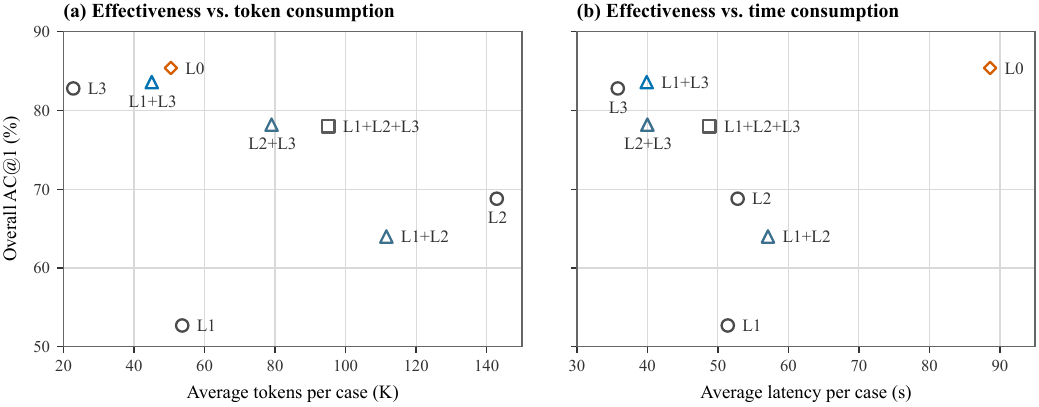}
    \caption{Effectiveness--efficiency trade-offs of tool configurations under Qwen3.7-Plus. Each point represents one configuration.}
    \label{fig:rq1-tradeoff}
\end{figure}

\textbf{Beyond L3, additional levels add cost with little return.}
Figure~\ref{fig:rq1-tradeoff} relates AC@1 to token and time consumption.
In both panels, only L3, L1+L3, and L0 lie on the effectiveness--efficiency frontier, and every other configuration is both less accurate and more costly than at least one of them.
L3 is the cheapest configuration, requiring 22.8K tokens and 35.8 seconds per case.
L1+L3 gains 0.8 points for roughly twice the tokens, and L0 gains 2.6 points at 2.5 times the latency, because the agent must write and execute its own analysis.
Adding L2 to L3 instead increases token consumption 3.5-fold while lowering AC@1.
For TA, L0 drops off the frontier, since L3 identifies failure types more accurately at lower cost.
Adding one level to L3 raises TA by 2.0 (L1) or 3.0 (L2) points, and the full toolset gains 3.4 points over L3 at 4.2 times the tokens.

\begin{finding}
For localization, a combination rarely outperforms its strongest constituent level, and only L3, L1+L3, and L0 lie on the effectiveness--efficiency frontier.
For type identification, every added level helps, but beyond L3 the gains are small relative to the cost.
\label{finding:composition}
\end{finding}

\begin{figure}[h!]
    \centering
    \includegraphics[width=1\textwidth]{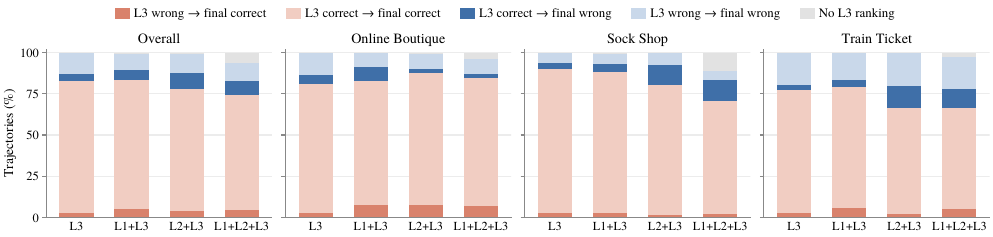}
    \caption{Correctness of the L3 candidate and the final diagnosis under Qwen3.7-Plus.
L3 and final indicate whether the top root cause candidate from the first L3 ranking and the agent's final diagnosis are correct, respectively.
No L3 ranking denotes trajectories that call no L3 tool returning ranked root cause candidates.}
    \label{fig:trace-outcome}
\end{figure}

\textbf{Why do combinations not add up for localization?}
We compare the four configurations with L3 and call the top candidate of the first L3 ranking in a trajectory its \emph{L3 candidate}.
Additional levels change how the investigation starts but not what L3 returns.
With L2+L3, the agent begins with anomaly filtering in 91\% of trajectories rather than with the L3 ranking.
Yet on the same case, L3 and L2+L3 return the same candidate in 99.9\% of trajectory pairs that obtain one, as do L1+L3 and L1+L2+L3.
This candidate is correct in more than 80\% of trajectories in every configuration (Figure~\ref{fig:trace-outcome}).
The accuracy differences therefore arise after L3 returns, when the agent decides whether to keep the candidate.

We call a trajectory a \emph{correction} when its L3 candidate is wrong but its final diagnosis is correct, and an \emph{override} in the opposite case.
Overrides outnumber corrections in every configuration, so adopting the L3 candidate directly would have been more accurate.
Additional levels shift this balance in opposite directions.
With L1+L3, the agent calls L1 after the L3 ranking in 98.5\% of trajectories, mostly to inspect the ranked candidates.
This verification doubles corrections from 2.7\% to 5.3\% of trajectories, while overrides rise only from 4.2\% to 5.8\%.
L2 instead supplies competing evidence, such as the metrics that its candidate filtering ranks as most anomalous, and the agent sometimes abandons the L3 candidate for the entity this evidence points to.
With L2+L3, overrides more than double to 9.3\%, while corrections rise only to 3.8\%; adding L2 to L1+L3 likewise raises overrides.

\begin{finding}
Additional levels change whether the agent keeps the L3 candidate rather than the candidate itself: L1 helps the agent verify it, whereas L2 competes with it and more often leads the agent away from a correct one.
\label{finding:decision}
\end{finding}

\textbf{Why does type identification remain difficult?}
The agent tends to name a fault after its most prominent symptom, and failure propagation makes CPU and latency anomalies prominent regardless of the injected fault.
Under L0, Qwen3.7-Plus identifies memory faults in only 28.9\% of trajectories and disk faults in none, labeling most of both as CPU faults; with L3 alone, disk faults are still identified in only 4.0\%.
The full toolset raises TA to 77.8\% for memory and 29.8\% for disk faults, but identifies delay faults less often than L0 (50.2\% vs.\ 84.4\%).
Packet loss, mostly labeled as delay or CPU faults, is identified in at most 0.9\% of trajectories in any of these configurations.
Type identification is thus not a by-product of localization: under L0, 46.2\% of all trajectories locate the correct service but name the wrong failure type.

\begin{finding}
Agents tend to name a fault after its most prominent symptom and thus mislabel memory and disk faults as CPU faults.
L1+L2+L3 reduces but does not remove this error, and packet loss is almost never identified.
\label{finding:type}
\end{finding}

\begin{rqanswer}{1}
Higher abstraction improves both localization and type identification, whereas combining levels rarely improves localization and, beyond L3, improves type identification only slightly.
Because additional levels affect whether the agent keeps the L3 candidate rather than the candidate itself, L3 alone is the most cost-effective choice for localization.
\end{rqanswer}

\subsection{RQ2: Does model capability affect the benefits of different tool levels?}
\label{sec:results:rq2}

RQ1 used the strongest model in our study.
A higher-level tool, however, performs more of the analysis, which helps only if the model can call the tool and use its output.
RQ2 therefore asks whether model capability affects the benefits of different tool levels.
We compare the four models along the path from L0 to the full structured toolset, namely L0, L1, L1+L2, and L1+L2+L3 (Table~\ref{tab:rq2_results}).

\begin{table}[hbpt]
\centering
\caption{Overall effectiveness and efficiency of different tool configurations across four Qwen models.
AC@1, AC@3, Avg@5, TA, and Completion are reported as percentages.
Tokens are reported in thousands, latency in seconds, and API cost in US dollars, all averaged per case.
The best value within each seed LLM is shown in bold.}
\label{tab:rq2_results}
\resizebox{\columnwidth}{!}{
\begin{tabular}{ll*{8}{c}}
\toprule
\multirow{2}{*}{Seed LLM}
& \multirow{2}{*}{Configuration}
& \multicolumn{5}{c}{Effectiveness}
& \multicolumn{3}{c}{Efficiency} \\
\cmidrule(lr){3-7}
\cmidrule(lr){8-10}
& & AC@1 & AC@3 & Avg@5 & TA & Completion
& Tokens & Latency & Cost \\
\midrule
\multirow{4}{*}{Qwen3-8B}
& L0       & 14.8 & 32.7 & 31.4 & 17.5 & 77.7  & 80.9  & 83.4 & 0.007 \\
& L1       & 29.3 & 48.4 & 47.1 & 24.4 & 96.4  & \textbf{63.0} & \textbf{41.8} & \textbf{0.005} \\
& L1+L2    & \textbf{59.6} & \textbf{77.2} & \textbf{74.2} & \textbf{28.7} & \textbf{98.8} & 90.7 & 46.5 & 0.007 \\
& L1+L2+L3 & 45.7 & 64.2 & 63.2 & 25.7 & 97.6  & 122.8 & 53.7 & 0.009 \\
\midrule
\multirow{4}{*}{Qwen3-32B}
& L0       & 39.6 & 57.6 & 56.1 & 24.4 & \textbf{94.0} & \textbf{53.5} & 75.3 & \textbf{0.018} \\
& L1       & 45.2 & 68.3 & 64.4 & 23.4 & 93.1  & 68.6  & \textbf{43.2} & 0.022 \\
& L1+L2    & 49.9 & 67.3 & 64.5 & 28.3 & 86.8  & 160.5 & 62.0 & 0.049 \\
& L1+L2+L3 & \textbf{54.0} & \textbf{72.4} & \textbf{69.9} & \textbf{31.1} & 87.7 & 193.8 & 66.2 & 0.059 \\
\midrule
\multirow{4}{*}{Qwen3-235B-A22B}
& L0       & 35.6 & 58.6 & 56.0 & 23.5 & 98.8  & 48.9  & 63.4 & 0.017 \\
& L1       & 45.0 & 66.8 & 63.7 & 24.3 & 97.5  & \textbf{47.2} & 40.5 & \textbf{0.015} \\
& L1+L2    & 63.9 & 83.8 & 80.7 & 39.3 & 98.8  & 78.2  & 38.8 & 0.024 \\
& L1+L2+L3 & \textbf{69.2} & \textbf{86.0} & \textbf{83.1} & \textbf{41.2} & \textbf{99.4} & 87.3 & \textbf{37.6} & 0.027 \\
\midrule
\multirow{4}{*}{Qwen3.7-Plus}
& L0       & \textbf{85.4} & 93.6 & \textbf{92.7} & 41.4 & 100.0 & \textbf{50.4} & 88.6 & 0.019 \\
& L1       & 52.7 & 70.0 & 69.1 & 24.3 & 100.0 & 53.7  & 51.4 & \textbf{0.018} \\
& L1+L2    & 64.0 & 88.4 & 84.4 & 47.3 & 100.0 & 111.6 & 57.1 & 0.035 \\
& L1+L2+L3 & 78.0 & \textbf{94.8} & 92.1 & \textbf{49.0} & 100.0 & 95.1 & \textbf{48.8} & 0.030 \\
\bottomrule
\end{tabular}
}
\end{table}

\begin{figure}[ht]
    \centering
    \includegraphics[width=0.95\textwidth]{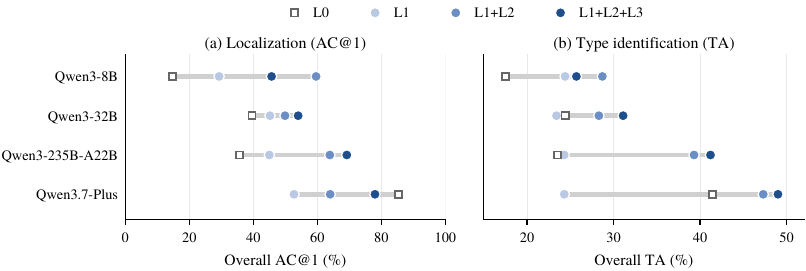}
    \caption{Overall AC@1 (a) and TA (b) of each tool configuration across the four Qwen models. Lines span the lowest and highest values for each model.}
    \label{fig:rq2-l0}
\end{figure}

\textbf{Predefined tools help weaker models localize, and evidence analysis helps every model identify failure types.}
Whether predefined tools help localization depends on the model (Figure~\ref{fig:rq2-l0}(a)).
For the three weaker models, every structured configuration achieves higher AC@1 than L0, and the best one exceeds L0 by 14.4 to 44.8 points.
For Qwen3.7-Plus, the order reverses: L0 outperforms all three.
For type identification, L1+L2 and L1+L2+L3 achieve higher TA than L0 for all four models, including Qwen3.7-Plus (49.0\% with L1+L2+L3 vs.\ 41.4\% with L0; Figure~\ref{fig:rq2-l0}(b)).
L1 alone, by contrast, raises TA substantially only for Qwen3-8B (+6.9 points) and lowers it by 17.1 points for Qwen3.7-Plus.

\begin{finding}
For localization, structured configurations outperform L0 for the three weaker models but not for Qwen3.7-Plus.
For type identification, configurations with L2 outperform L0 for all four models.
\label{finding:l0}
\end{finding}

\begin{figure}[htbp]
    \centering
    \includegraphics[width=1\textwidth]{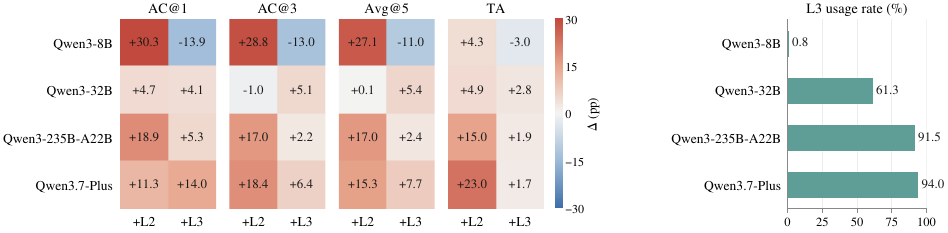}
    \caption{
\textbf{Left:} Change in overall effectiveness after adding L2 (L1$\rightarrow$L1+L2) and L3 (L1+L2$\rightarrow$L1+L2+L3). \textbf{Right:} L3 usage rate under L1+L2+L3.
}
    \label{fig:rq2-heat-map}
\end{figure}

\textbf {L2 helps every model, whereas adding L3 helps all but the weakest.}
Adding L2 to L1 improves AC@1 for all four models, by 4.7 to 30.3 points, though not in order of capability (Figure~\ref{fig:rq2-heat-map}, left).
Its TA gain, by contrast, grows with capability, from 4.3 points for Qwen3-8B to 4.9, 15.0, and 23.0 points for the larger models.
Adding L3 splits the models: Qwen3-8B declines on all four effectiveness metrics, losing 13.9 points in AC@1, whereas the other three improve on all four, although their TA gains stay below 3 points.
L3 usage mirrors this split (Figure~\ref{fig:rq2-heat-map}, right): under L1+L2+L3, Qwen3-8B calls L3 in 0.8\% of trajectories, compared with 61.3\%, 91.5\%, and 94.0\% for the other three models.
The two smaller models also gain little in type identification overall: their TA never exceeds 31.1\%, only moderately above the 20\% expected from guessing among five balanced failure types, even where their AC@1 reaches 54.0--59.6\%.

\begin{finding}
Adding L2 improves both targets for all four models, with TA gains that grow with capability.
Adding L3 helps the three models that call it and hurts Qwen3-8B, which rarely does.\label{finding:marginal}
\end{finding}

\begin{figure}[hpbt]
    \centering
    \includegraphics[width=1\textwidth]{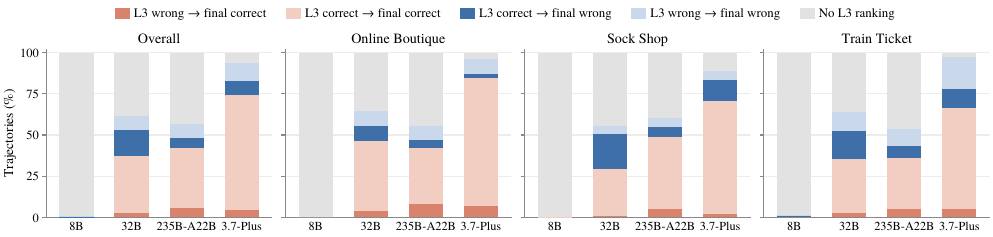}
    \caption{Correctness of the L3 candidate and the final diagnosis across models under L1+L2+L3.
L3 and final indicate whether the top root cause candidate from the first L3 ranking and the agent's final diagnosis are correct, respectively.
No L3 ranking denotes trajectories that call no L3 tool returning ranked root cause candidates.}
    \label{fig:trace-outcome-models}
\end{figure}

\textbf{Why does L3 help some models but not others?}
A tool level improves the diagnosis insofar as the agent delegates analysis to it, calls it properly, and keeps its correct results.
Figure~\ref{fig:trace-outcome-models} shows that the models fall short at different points.
Qwen3-8B falls short at delegation, obtaining an L3 ranking in almost no trajectory.
Like the other Qwen3 models, it often searches on its own before any tool proposes a candidate, inspecting services one by one from the entry service (43--78\% of trajectories across the three Qwen3 models), which requires only metric access.
Qwen3.7-Plus never does so and instead starts by filtering candidate metrics across the whole case.
The decline of Qwen3-8B after L3 is added therefore does not come from L3 outputs.
Instead, the larger toolset disrupts its own search: its trajectories grow from 10.1 to 11.7 steps on average and contain more invalid actions, such as malformed tool calls, and the decline concentrates on Sock Shop, where its AC@1 drops by 33.9 points.

The models that do call L3 fall short at later points.
Qwen3-235B-A22B obtains an L3 ranking in only about half of its trajectories, and in about a third of its evidence fusion calls it narrows the analysis window itself, which lowers the accuracy of the L3 candidate from 83\% to 53\%.
Qwen3-32B obtains L3 candidates as accurate as those of Qwen3.7-Plus but overrides nearly a third of the correct ones, losing the most at the decision; its trajectories are also the costliest in Table~\ref{tab:rq2_results}, at 193.8K tokens per case.
Qwen3.7-Plus delegates in nearly every trajectory, and, as shown in RQ1, its losses concentrate on overrides.
The models also rank differently on different abilities: Qwen3-32B outperforms Qwen3-235B-A22B under L0 (39.6\% vs.\ 35.6\%), yet it handles L3 results less reliably.

\begin{finding}
Models lose accuracy at different points of delegation: Qwen3-8B rarely delegates to L3, Qwen3-235B-A22B often narrows its time window when calling it, and Qwen3-32B frequently overrides its correct results.
\label{finding:delegation}
\end{finding}

\begin{rqanswer}{2}
Predefined tools help weaker models localize, and evidence analysis helps every model identify failure types.
How much higher-level tools pay off depends on whether a model delegates to them, calls them properly, and keeps their correct results; for models that rarely call L3, L1+L2 is the better configuration.
\end{rqanswer}

\subsection{RQ3: Does system complexity affect the effectiveness of different tool configurations?}
\label{sec:results:rq3}

\begin{figure}[h!]
    \centering
    \includegraphics[width=1\textwidth]{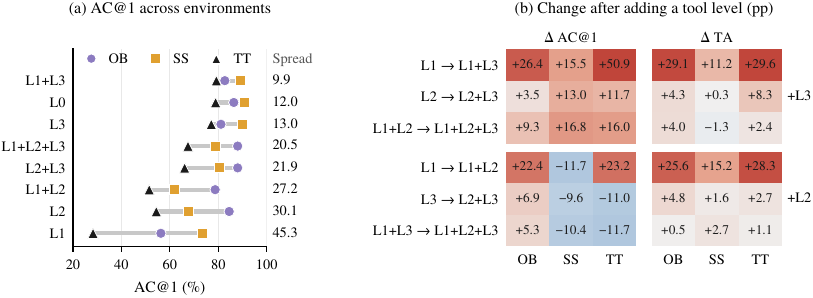}
    \caption{
Cross-system effectiveness of tool configurations under Qwen3.7-Plus. (a) AC@1 in each environment, with configurations sorted by their cross-environment spread (max$-$min). (b) Change in AC@1 (left) and TA (right) after adding L3 (top) or L2 (bottom) to a base configuration, in percentage points; red marks a gain and blue a loss. OB, SS, and TT have 12, 15, and 64 services, respectively.
}

    \label{fig:rq3-heat-map}
\end{figure}

RQ1 and RQ2 pool results over the three systems, although RQ1 already shows that the ordering of tool levels differs on Sock Shop.
The systems share the same protocol, number of cases, and failure types, but differ in service scale, communication structure, and metric dimensionality (Table~\ref{tab:diagnostic-environments}).
We analyze Qwen3.7-Plus, which uses all tool levels (RQ2), and use the other models to check whether its patterns generalize.
Tables~\ref{tab:rq3_config_system} and~\ref{tab:rq3_model_system} report per-system results for all configurations under Qwen3.7-Plus and for the four models, respectively.
Figure~\ref{fig:rq3-heat-map} shows AC@1 in each system and the effect of adding L2 or L3 on AC@1 and TA.

\begin{table}[bhpt]
\centering
\caption{Per-system effectiveness of different tool configurations using Qwen3.7-Plus.
AC@1, AC@3, Avg@5, and TA are reported as percentages.
The best value(s) in each column are shown in bold; ties are all bolded.}
\label{tab:rq3_config_system}
\resizebox{\textwidth}{!}{
\begin{tabular}{ll*{12}{c}}
\toprule
\multirow{2}{*}{Seed LLM}
& \multirow{2}{*}{Configuration}
& \multicolumn{4}{c}{Online Boutique}
& \multicolumn{4}{c}{Sock Shop}
& \multicolumn{4}{c}{Train Ticket} \\
\cmidrule(lr){3-6}
\cmidrule(lr){7-10}
\cmidrule(lr){11-14}
& & AC@1 & AC@3 & Avg@5 & TA
& AC@1 & AC@3 & Avg@5 & TA
& AC@1 & AC@3 & Avg@5 & TA \\
\midrule
\multirow{8}{*}{Qwen3.7-Plus}
& L0       & 86.4 & 95.2 & 94.0 & 45.9 & \textbf{90.9} & 96.3 & 95.9 & 41.6 & 78.9 & 89.3 & 88.2 & 36.8 \\
& L1       & 56.3 & 76.0 & 73.7 & 23.2 & 73.6 & 87.5 & 86.5 & 28.8 & 28.3 & 46.7 & 47.1 & 20.8 \\
& L2       & 84.5 & 95.2 & 93.2 & 50.1 & 67.5 & 98.4 & 92.0 & 40.8 & 54.4 & 83.2 & 77.2 & 42.1 \\
& L3       & 81.1 & 96.0 & 93.5 & 49.6 & 90.1 & \textbf{100.0} & \textbf{97.8} & 39.5 & 77.1 & \textbf{92.5} & \textbf{89.8} & 47.7 \\
& L1+L2    & 78.7 & 91.5 & 89.2 & 48.8 & 61.9 & 96.8 & 90.1 & \textbf{44.0} & 51.5 & 76.8 & 73.8 & 49.1 \\
& L1+L3    & 82.7 & 93.6 & 92.0 & 52.3 & 89.1 & 98.1 & 96.0 & 40.0 & \textbf{79.2} & 91.2 & 89.4 & 50.4 \\
& L2+L3    & \textbf{88.0} & \textbf{97.1} & \textbf{95.6} & \textbf{54.4} & 80.5 & 99.2 & 95.0 & 41.1 & 66.1 & 90.4 & 86.2 & 50.4 \\
& L1+L2+L3 & \textbf{88.0} & 96.3 & 95.2 & 52.8 & 78.7 & 98.9 & 94.8 & 42.7 & 67.5 & 89.3 & 86.1 & \textbf{51.5} \\
\bottomrule
\end{tabular}
}
\end{table}

\textbf{Train Ticket is the hardest system for localization, but not for type identification.}
Every configuration achieves its lowest AC@1 on Train Ticket, which has 64 services and 198--238 metrics per case, compared with 12--15 services and 49--63 metrics for the other two systems.
The degradation, however, is far from uniform (Figure~\ref{fig:rq3-heat-map}(a)).
The cross-system spread of AC@1 narrows as abstraction rises, from 45.3 points for L1 to 30.1 for L2 and 13.0 for L3, and L1+L3 and L0 are the most stable configurations, with spreads of 9.9 and 12.0 points.
Trajectories show that the large system mainly hurts the agent's own search.
With L1 alone, the agent never considers the true root cause in 30.1\% of Train Ticket trajectories, compared with at most 2.4\% on the other two systems.
Adding L3 narrows the search, reducing this share to 4.3\% and raising AC@1 on Train Ticket by 50.9 points, the largest gain in Figure~\ref{fig:rq3-heat-map}(b).
Type identification is also hardest on Train Ticket without analysis tools, with TA of 36.8\% under L0 and 20.8\% under L1 (Table~\ref{tab:rq3_config_system}).
Once L3 is available, however, Train Ticket comes within 4.0 points of Online Boutique in TA, and Sock Shop becomes the hardest system, where no configuration raises TA more than 2.4 points above L0.

\begin{finding}
Every configuration localizes worst on Train Ticket, the largest system, and low-abstraction configurations suffer most: the larger system mainly hurts the agent's own search, which L3 compensates for.
For type identification, Train Ticket is hardest only without analysis tools; with them, Sock Shop is, and tools barely raise its TA.
\label{finding:scale}
\end{finding}

\begin{table}[h!]
\centering
\caption{Per-system effectiveness of different tool configurations across four Qwen models.
AC@1, AC@3, Avg@5, and TA are reported as percentages.
The best value(s) within each seed LLM are shown in bold; ties are all bolded.}
\label{tab:rq3_model_system}
\resizebox{\textwidth}{!}{
\begin{tabular}{ll*{12}{c}}
\toprule
\multirow{2}{*}{Seed LLM}
& \multirow{2}{*}{Configuration}
& \multicolumn{4}{c}{Online Boutique}
& \multicolumn{4}{c}{Sock Shop}
& \multicolumn{4}{c}{Train Ticket} \\
\cmidrule(lr){3-6}
\cmidrule(lr){7-10}
\cmidrule(lr){11-14}
& & AC@1 & AC@3 & Avg@5 & TA
& AC@1 & AC@3 & Avg@5 & TA
& AC@1 & AC@3 & Avg@5 & TA \\
\midrule
\multirow{4}{*}{Qwen3-8B}
& L0       & 16.5 & 37.3 & 35.8 & 20.0 & 14.7 & 34.9 & 31.5 & 13.1 & 13.1 & 25.9 & 26.8 & 19.5 \\
& L1       & 27.5 & 55.7 & 51.9 & 26.4 & 38.1 & 57.1 & 56.1 & 23.2 & 22.4 & 32.5 & 33.2 & 23.7 \\
& L1+L2    & \textbf{59.7} & \textbf{85.1} & \textbf{79.7} & \textbf{32.0} & \textbf{74.7} & \textbf{89.3} & \textbf{87.0} & \textbf{26.4} & \textbf{44.5} & 57.1 & 56.0 & 27.7 \\
& L1+L2+L3 & 53.1 & 78.9 & 76.3 & 27.5 & 40.8 & 54.9 & 56.5 & 21.1 & 43.2 & \textbf{58.7} & \textbf{56.9} & \textbf{28.5} \\
\midrule
\multirow{4}{*}{Qwen3-32B}
& L0       & 45.6 & 60.8 & 60.5 & 26.1 & 45.9 & 69.6 & 66.2 & \textbf{26.1} & 27.2 & 42.4 & 41.8 & 21.1 \\
& L1       & 48.3 & 72.5 & 68.4 & 24.3 & 54.1 & \textbf{76.0} & \textbf{72.3} & 23.2 & 33.3 & 56.3 & 52.6 & 22.7 \\
& L1+L2    & \textbf{65.1} & \textbf{82.7} & \textbf{79.6} & \textbf{36.8} & \textbf{60.3} & 74.1 & 71.9 & 25.3 & 24.3 & 45.1 & 41.9 & 22.7 \\
& L1+L2+L3 & 62.7 & 81.1 & 79.1 & 36.0 & 55.7 & 72.8 & 69.9 & 22.4 & \textbf{43.7} & \textbf{63.5} & \textbf{60.8} & \textbf{34.9} \\
\midrule
\multirow{4}{*}{Qwen3-235B-A22B}
& L0       & 42.1 & 65.9 & 62.5 & 25.1 & 40.0 & 64.8 & 62.7 & 21.9 & 24.8 & 45.1 & 42.8 & 23.5 \\
& L1       & 49.1 & 70.4 & 68.5 & 29.1 & 58.1 & 81.1 & 75.8 & 25.1 & 27.7 & 49.1 & 46.7 & 18.7 \\
& L1+L2    & 69.6 & 88.5 & 86.0 & 44.5 & 68.8 & 89.6 & 85.9 & 33.1 & 53.3 & \textbf{73.3} & 70.1 & 40.3 \\
& L1+L2+L3 & \textbf{75.7} & \textbf{93.3} & \textbf{89.7} & \textbf{48.3} & \textbf{74.7} & \textbf{91.2} & \textbf{88.3} & \textbf{34.1} & \textbf{57.3} & \textbf{73.3} & \textbf{71.4} & \textbf{41.1} \\
\midrule
\multirow{4}{*}{Qwen3.7-Plus}
& L0       & 86.4 & 95.2 & 94.0 & 45.9 & \textbf{90.9} & 96.3 & \textbf{95.9} & 41.6 & \textbf{78.9} & \textbf{89.3} & \textbf{88.2} & 36.8 \\
& L1       & 56.3 & 76.0 & 73.7 & 23.2 & 73.6 & 87.5 & 86.5 & 28.8 & 28.3 & 46.7 & 47.1 & 20.8 \\
& L1+L2    & 78.7 & 91.5 & 89.2 & 48.8 & 61.9 & 96.8 & 90.1 & \textbf{44.0} & 51.5 & 76.8 & 73.8 & 49.1 \\
& L1+L2+L3 & \textbf{88.0} & \textbf{96.3} & \textbf{95.2} & \textbf{52.8} & 78.7 & \textbf{98.9} & 94.8 & 42.7 & 67.5 & \textbf{89.3} & 86.1 & \textbf{51.5} \\
\bottomrule
\end{tabular}
}
\end{table}

\textbf{Adding L3 helps in every system, whereas the effect of L2 reverses for localization only.}
Figure~\ref{fig:rq3-heat-map}(b) contrasts the two levels.
Adding L3 improves AC@1 in all nine combinations of base configuration and system.
Qwen3-235B-A22B, which calls L3 in 91.5\% of trajectories, shows the same consistency, gaining 4.0 to 6.1 points from L3 in every system, unlike Qwen3-32B (Table~\ref{tab:rq3_model_system}).
The effect of adding L2 instead depends on the system.
It improves all three base configurations on Online Boutique and lowers all three on Sock Shop, by 9.6 to 11.7 points.
On Train Ticket, it improves L1 by 23.2 points but lowers the configurations that already include L3 by 11.0 and 11.7 points.
System scale alone cannot explain this reversal, because L2 hurts on Sock Shop, which is nearly as small as Online Boutique.
We observe the reversal for Qwen3.7-Plus; for the three weaker models, whose L1 accuracy is lower, adding L2 to L1 improves AC@1 on Sock Shop as well (Table~\ref{tab:rq3_model_system}).
The reversal is also specific to localization: as the right half of Figure~\ref{fig:rq3-heat-map}(b) shows, adding L2 raises TA in all nine combinations, including those on Sock Shop and Train Ticket where it lowers AC@1, and adding L3 raises TA in eight of nine.

\begin{finding}
Adding L3 improves localization in every system, whereas adding L2 reverses its effect on localization across systems, even between the two small ones, yet raises TA in every system.
\label{finding:l2-reversal}
\end{finding}

\textbf{Why does the effect of L2 reverse across systems?}
The per-system panels of Figure~\ref{fig:trace-outcome} show how L2 changes the agent's decision on the L3 candidate.
On Sock Shop and Train Ticket, adding L2 to L3 raises overrides from 3.5\% of trajectories to 12.0\% and 13.6\%, respectively.
On Online Boutique, it instead lowers overrides from 5.6\% to 2.4\% and nearly triples corrections, so that the agent's decision improves on the L3 candidate.
The difference lies in whether the entity that L2 ranks as most anomalous is the root cause.
When this entity differs from the L3 candidate and the agent switches to it, the switch is correct in 79\% of cases on Online Boutique but in only 5\% on Sock Shop and Train Ticket.
That L2 still raises TA on these systems suggests that its evidence often reveals what kind of fault occurred even when it points to the wrong entity.
Because service scale, communication structure, and metric composition vary jointly across our systems, we do not attribute this difference to any single system property.

\begin{finding}
For localization, L2 evidence helps the agent where the entity it ranks as most anomalous tends to be the root cause, and misleads the agent where that entity tends to be another component.
\label{finding:l2-evidence}
\end{finding}

\begin{rqanswer}{3}
A larger system mainly hurts the agent's own search, which L3 compensates for in every system.
L2 helps localization only where the most anomalous signals point to the root cause, yet improves TA everywhere; when a toolset moves to a new system, L2 is the level whose localization benefit needs re-validation.
\end{rqanswer}

\section{Discussion}
\label{sec:discussion}

\subsection{Failure Analysis}
\label{sec:discussion:failure}

RQ1--RQ3 identify the tool configurations that work best; we now ask why even the best of them fail.
With its most accurate structured toolset, L1+L3, Qwen3.7-Plus still selects a wrong root cause in 16.4\% of cases, and its TA never exceeds 49.0\%.
We examine its failed trajectories under the configurations with L2 or L3.

\textbf{The agent finds the root cause but does not keep it.}
Qwen3.7-Plus completes every case, considers the true root cause in at least 96\% of trajectories, and receives a correct top L3 candidate in more than 80\% of the trajectories with an L3 ranking.
Yet the agent had considered the true root cause in 84--95\% of trajectories with incorrect localizations, indicating that localization failures arise mainly during the final decision rather than the search.
Two patterns recur.
The agent mistakes a service affected by the fault for its source, which is how L2 leads it away from correct L3 candidates on Sock Shop and Train Ticket (RQ1, RQ3).
In a Train Ticket case with a delay fault in \texttt{ts-auth-service}, L3 ranked the correct service first, but candidate filtering ranked a memory metric of the payment service highest; finding no CPU or memory anomaly in the auth service, the agent treated its latency as a propagated symptom and blamed the payment service.
The agent also uses tools to confirm rather than test a hypothesis: the evidence chain interface gathers observations that support a given candidate, and when that candidate is wrong, the agent keeps it in 83\% of such calls.

\textbf{The agent names the fault after its most prominent symptom.}
Type identification fails more often than localization: when Qwen3.7-Plus locates the root cause, it names the failure type correctly in only 55.7\% of such trajectories under L1+L2+L3 and 45.9\% under L0.
The agent infers the failure type from anomalous metrics highlighted in the tool outputs.
Since failure propagation makes CPU and latency anomalies prominent whatever the injected fault, these metrics point to the symptom rather than the fault: even with L3, disk faults are identified in only 4.0\% of trajectories, and packet loss is almost never identified (RQ1).
More evidence shifts these errors rather than removing them, as the full toolset identifies memory and disk faults more often but delay faults less often than L0 (50.2\% vs.\ 84.4\%).

\textbf{Both failures call for tools that discriminate between explanations.}
In both cases, the agent must choose between explanations that the evidence fits equally well: the source of a fault versus the services it propagates to, and one failure type versus another with the same symptoms.
Tools that rank evidence by how anomalous it is leave this choice unresolved, consistent with the small gains from adding levels beyond L3 (RQ1).
For capable models, tools that seek counter-evidence or rank evidence by how well it separates failure types are thus more promising than tools that add analysis.
Weaker models fail earlier: they leave more cases unfinished (Table~\ref{tab:rq2_results}) and, under L0, never consider the root cause in 41--51\% of trajectories.
They benefit first from a small and robust tool interface.

\subsection{Threats to Validity}
\label{sec:discussion:validity}

\textbf{Internal validity.}
Our implemented toolset covers representative RCA capabilities across abstraction levels, and each function was unit-tested to verify its intended behavior.
While these tools support the comparisons in our study, they do not cover the full range of RCA tools within each abstraction level.
We also compare tool abstraction levels and their combinations within a common agent harness, allowing the agent to select individual interfaces during diagnosis.
The results therefore characterize the overall effects of the evaluated tool configurations, while broader tool coverage and finer-grained attribution to specific interfaces within each level remain directions for further investigation.

\textbf{External validity.}
We evaluate all 375 RE1 cases across three microservice systems and five injected failure types, providing coverage across different benchmark environments.
This coverage supports comparisons within the benchmark, but production incidents and diagnosis incorporating logs and traces remain outside our experimental scope.
In addition, our model comparison covers four models from the Qwen lineage, allowing us to examine capability differences within a common family.
However, the applicability of these findings beyond the Qwen family has not yet been assessed.
Extending this comparison to other model families therefore remains a direction for future study.
\section{Related Work}
\label{sec:related-work}

\textbf{Agentic RCA for microservice systems.}
Agentic RCA systems adopt different strategies for providing analytical support and organizing tool use.
Preprocessing-based approaches encapsulate operations such as telemetry filtering, anomaly detection, and candidate scoring in predefined pipelines, allowing agents to reason over the resulting evidence~\citep{luo2025opsagent,sun2025trioxpert,tang2025microrcaagent}.
Code-based approaches provide a general execution interface through which agents implement data retrieval and analysis, as exemplified by OpenRCA's RCA-agent and reflected in our L0 setting~\citep{xu2025openrca}.
Workflow-guided approaches use SOPs and structured operational knowledge to organize tool invocation and code generation~\citep{pei2025flowofaction,gao2026rcaflow}.
Reasoning-oriented approaches employ agent collaboration, search, explicit states, or memory to guide evidence acquisition and the interpretation of tool results~\citep{zhang-etal-2024-mabc,naakka2026latsrca,luo2026graph,zhang2026agentic,sun2026opsmem, luo2026memoryharnessinglonghorizonagents}.
However, these studies primarily evaluate tool support within specific diagnostic frameworks, providing limited guidance on how tool abstraction and composition affect diagnosis.

\textbf{Empirical studies of RCA and tool use.}
RCA benchmarks provide the experimental basis for comparing diagnostic methods across microservice systems, telemetry sources, and fault propagation scenarios~\citep{pham2025rcaeval,xu2025openrca,fang2026rethinking}.
Beyond RCA, tool-use benchmarks assess whether models can retrieve appropriate tools, invoke APIs correctly, and complete multi-step interactions~\citep{li2023apibank,qin2024toolllm,lu-etal-2025-toolsandbox,pmlr-v267-patil25a}.
Controlled studies address tool design more directly, comparing interface organization under similar capabilities in coding agents and examining tool complexity and composition in web agents~\citep{xu2026toolarchitecture,lou2026toolillusion}.
Extending these evaluations to the division of analytical work in RCA, our study compares the diagnostic value of tools that provide observations, analytical evidence, or root cause recommendations, individually and in combination.
\section{Conclusion}
\label{sec:conclusion}

We conducted a controlled empirical study of tool abstraction and composition for agentic RCA across different models and microservice systems.
Our results show that tool configurations affect root cause localization and failure type identification differently.
With the strongest evaluated model, L3 diagnostic tools offer a favorable balance between localization accuracy and cost.
Adding tool levels improves overall failure type identification, but can reduce localization accuracy.
These benefits also depend on the model: adding L3 to L1+L2 improves localization for three of the four models, but reduces it for the weakest model.
Even with the same model, the relative performance of tool configurations changes across microservice systems.
Together, these findings provide an empirical foundation for designing and using RCA tools, guiding practitioners to select tool abstraction and composition according to the diagnostic objective, model capability, and target system.

\clearpage
\bibliographystyle{plainnat}
\bibliography{references}

\end{document}